\documentclass[conference]{IEEEtran}
\usepackage{cite,graphicx,amssymb,amsmath,color,textcomp}
\usepackage[square, comma, sort&compress, numbers]{natbib}

\usepackage{algorithm}
\usepackage{algorithmicx}
\usepackage{graphicx}
\usepackage{epsfig}
\usepackage{epstopdf}
\usepackage{mathrsfs}
\usepackage{color}
\usepackage{bm}
\usepackage{epstopdf}
\usepackage{amsfonts,amssymb}
\usepackage{amsthm}
\usepackage{setspace}
\usepackage{amsmath}
\usepackage{algpseudocode}
\usepackage{amssymb}

\begin{document}

\title{Optimal Design of SWIPT-Aware Fog Computing Networks}

\author{Jingxian Liu$^{*,\Diamond}$, Ke Xiong$^{*,\Diamond}$, Pingyi Fan$^\dag$, Zhangdui Zhong$^{\natural,\S}$ and Khaled Ben Letaief$^\ddag$\\
\small
$^*$School of Computer and Information Technology, Beijing Jiaotong University, Beijing 100044, China \\
$^\Diamond$ Beijing Key Laboratory of Traffic Data Analysis and Mining, Beijing Jiaotong University, Beijing 100044, China\\
$^\dag$Department of Electronic Engineering, Tsinghua University,
Beijing 100084, China\\
$^\natural$ State Key Lab of Rail Traffic Control and Safety, Beijing Jiaotong University, Beijing 100044, China\\
$^\S$Beijing Engineering Research Center of High-speed Railway Broadband Mobile Communications,\\
Beijing Jiaotong University, Beijing 100044, China \\
$^\ddag$ The Hong Kong University of Science and Technology, Hong Kong\\
Email: kxiong@bjtu.edu.cn
 }

\maketitle
\begin{abstract}
This paper studies a simultaneous wireless information and power transfer (SWIPT)-aware fog computing network, where a \textcolor[rgb]{0.00,0.00,0.00}{multiple antenna} fog function integrated hybrid access point (F-HAP) transfers information and energy to multiple heterogeneous single-antenna sensors and also helps some of them fulfill computing tasks. By jointly optimizing energy and information beamforming designs at the F-HAP, the bandwidth allocation and the computation offloading distribution, an optimization problem is formulated to minimize the required energy under communication and computation requirements, as well as energy harvesting constraints. Two  optimal designs, i.e., fixed offloading time (FOT) and optimized offloading time (OOT) designs, are  \textcolor[rgb]{0.00,0.00,0.00}{proposed}. As both designs get involved in solving non-convex problems, there are no known solutions to them. Therefore, for the FOT design, the semidefinite relaxation (SDR) is adopted to solve it. It is theoretically proved that the rank-one constraints are always satisfied, so the global optimal solution is guaranteed. For the OOT design, since its non-convexity is hard to deal with, a penalty dual decomposition (PDD)-based algorithm is proposed, which is able to achieve a suboptimal solution. The computational complexity for two designs are analyzed. Numerical results show that the partial offloading mode is superior to binary benchmark modes. \textcolor[rgb]{0.00,0.00,0.00}{It is also shown that if the system is with strong enough computing capability, the OOT design is suggested to achieve lower required energy; Otherwise, the FOT design is preferred to achieve a relatively low computation complexity.}
\end{abstract}


\section{Introduction}

Recently, to sustainably power wireless devices (WDs) in lower-power Internet of Things (IoT) systems, e.g., wireless sensor networks \cite{XB}, \textcolor[rgb]{0.00,0.00,0.00}{radio frequency (RF)-based energy harvesting (EH), capable of harvesting energy from stable and controllable RF signals, is envisioned as a promising solution.
One of the most popular application paradigms of RF EH is wireless powered communication networks (WPCNs) \cite{XD}, with which WDs firstly harvest energy from RF signals and then fulfill the communication and computing operations with the harvested energy. Since RF signals carry both information and energy, simultaneous wireless information and power transfer (SWIPT) was proposed as another popular RF EH application paradigm. It was reported that by integrating beamforming technology, heterogeneous devices (e.g., information decoding (ID) and energy harvesting (EH) devices) could be served by SWIPT to meet their different requirements \cite{Y. Lu0}.}

On the other hand, to enhance the computing capacity of IoT systems, fog computing (or mobile edge computing (MEC) as the alternative\footnote{Since some works considered mobile secarios, e.g., \cite{XC}.}) was proposed, which is able to reduce the long transmission delay by pushing computing, network control and storage functionalities to the network edge \cite{M. Chiang}. With fog computing, IoT WDs may offload part or all of the computation tasks to the fog server located at the network edge. There are two modes for IoT WDs to fulfill computation offloading, i.e., partial offloading and binary offloading. The partial offloading mode suits for the divisible computation tasks, which allows the computation task to be divided into two parts and  \textcolor[rgb]{0.00,0.00,0.00}{among them one part is offloaded to the fog server. }The binary offloading mode suits for the indivisible computation tasks, which requires the whole computation task to be either offloaded or locally computed.

To inherent the benefits of both RF EH and fog computing, some recent works have started to integrate them in a single IoT systems, see e.g., \cite{C. You}-\cite{S. Bi}. However, these works only considered WPCN systems with fog computing, and no SWIPT was involved. Since SWIPT realizes wireless power transfer (WPT) and wireless information transfer (WIT) at the same time, which is more suitable for latency-sensitive applications. A few recent works began to study SWIPT-aware fog computing systems. In \cite{N. Janatian} and \cite{H.N. Zheng}, the energy consumption was minimized by adopting time switching (TS) mode and power splitting (PS) mode, respectively, where a user was powered by an energy access point and then the tasks were offloaded to a fog server. In \cite{H. Chai}, the energy consumption was minimized by optimizing power, time and data allocation, \textcolor[rgb]{0.00,0.00,0.00}{where multiple users were considered.}

In this paper, we also focus on the optimal design of the SWIPT-aware fog computing system, where the \textcolor[rgb]{0.00,0.00,1.00}{fog function integrated hybrid access point (F-HAP)} first transmits energy and information to EH and ID devices with SWIPT. Then EH devices complete the computation tasks with the harvested energy and fog computing paradigm. Compared with existing works, the following differences should be emphasized.

{\it Firstly}, in existing works, see e.g. \cite{N. Janatian}-\cite{H.N. Zheng}, only single type of users or single user were studied. \textcolor[rgb]{0.00,0.00,0.00}{That is, only EH users, PS users or TS users were considered in their works.} In view of that in practice, it is very common to deploy various types of WDs with different EH or ID requirements in a single system, we consider heterogeneous users in our work, where both EH and ID devices are investigated.

 {\it Secondly}, in existing works, although beamforming vector was optimally designed to enhance transmission efficiency for SWIPT-enable fog computing networks, see e.g. \cite{H.N. Zheng}, energy and information signals were transmitted with the same beam vector. Considering that information and energy transmissions are with very different physical features and receiving sensitivities (e.g., $-60$ dBm for ID receivers and $-10$ dBm for EH receivers), \textcolor[rgb]{0.00,0.00,0.00}{we design different beamforming vectors and matrix for them}. By doing so, more flexibility is achieved, which therefore is able to yield a better system performance.

 {\it Thirdly}, in existing works, only a part of the following system resources and configurations, i.e., transmit beamforming vectors and matrix, bandwidth allocation, time assignment, and computation offloading distribution, were jointly optimized, see e.g., \cite{N. Janatian}-\cite{H. Chai}. In our work, all the configurations and resources mentioned above are jointly optimized in order to achieve the higher system performances.


The contributions of our work are summarized as follows.
\begin{itemize}
\item An optimization problem is formulated to minimize the F-HAP's required energy, where two optimal designs, i.e., the fixed offloading time (FOT) and the optimized offloading time (OOT) designs, are \textcolor[rgb]{0.00,0.00,0.00}{ presented}. In the first one, energy and information beamforming designs, bandwidth allocation and computation offloading distribution, are jointly optimized. In the second one, the offloading time assignment is also jointly optimized with the system configurations and resources mentioned in the FOT design.
\item For the FOT design, since the primal problem is non-convex, it is relaxed by adopting semidefinite relaxation (SDR) and solved by convex optimization problem solution methods, and then, we theoretically prove that the rank-one constraints are always satisfied. So, the global optimal solution is achieved. In order to provide more insights, a semi-closed optimal solution is also presented by using the dual decomposition method.
\item For the OOT design, since the problem is more complex and cannot be solved by using the solution method for the FOT design, we therefore present a penalty dual decomposition (PDD)-based algorithm, which is able to find the suboptimal solution.
\item  \textcolor[rgb]{0.00,0.00,0.00}{Based on the theoretical analysis and simulation results, we also discuss the performance of both proposed methods to solve both designs in terms of required energy and system computing capability (computational complexity). It suggests that if the system is with strong computing capability, the OOT design is suggested to achieve lower required energy; Otherwise, the FOT design is preferred to achieve a relatively low computation complexity.}
\end{itemize}

The rest of this paper is organized as follows. Section II describes the system model. Section III and Section IV present two optimal designs, respectively. Section V analyzes the computational complexity. Section VI shows some numerical results and Section VII concludes this paper.

\begin{figure}[!t]
\centering
\includegraphics[width=0.45\textwidth]{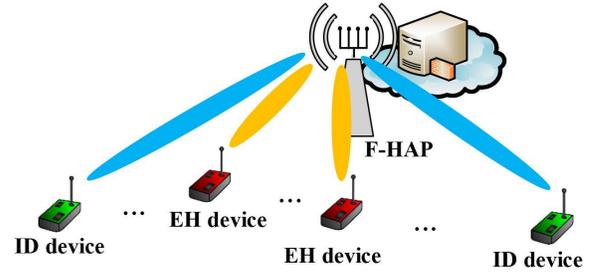}
\caption{Illustration of the SWIPT-aware fog computing system with a F-HAP and heterogeneous devices.}
\label{fig1}
\vspace{-3mm}
\end{figure}

%

\section{System Model}
Consider a SWIPT-aware fog computing system with a $N_t$-antenna F-HAP and heterogeneous single-antenna IoT devices as shown in Fig. \ref{fig1},
 where the F-HAP is deployed \textcolor[rgb]{0.00,0.00,0.00}{to transmit energy and information, as well as providing computation services}. Both EH and ID devices exist in the system, where the set of EH devices is denoted as $\mathsf{N}^{\rm {eh}} \buildrel \Delta \over =  \{1,..., {N}^{\rm {eh}}\}$ and that of ID devices is denoted as $\mathsf{N}^ {\rm {id}}\buildrel \Delta \over =  \{1,..., {N}^{\rm {id}}\}$. \textcolor[rgb]{0.00,0.00,0.00}{ID devices desire to receive information from the F-HAP, while EH devices desire to harvest energy from it.}
To enhance WPT and WIT efficiency, beamforming technology is employed at the F-HAP, and the energy and information are transmitted simultaneously via different beam vectors. \textcolor[rgb]{0.00,0.00,0.00}{Denote $\bm{\mathrm{h}}_{m,n}^{(\rm eh)}  \in {\mathbb{C}^{{N_t}\times1}}$ and $\bm{\mathrm{h}}_{m,n}^{(\rm id)}  \in {\mathbb{C}^{{N_t}\times1}}$ as the channel coefficients between node $m$ and node $n$ associated with the EH and ID devices, respectively\footnote{We assume that node $0$ represents the F-HAP.}.} The channel coefficients remain constants over time $T$ since the block fading channel model is assumed.

\subsection{DL Model}
In the DL, the F-HAP transmits energy and information to EH and ID devices at the same time, where the transmitted symbol is expressed by
$\bm{\mathrm{x}} = \sum\nolimits_{i = 1}^{{N^{\rm {eh}}}} {\bm{\mathrm{u}}_is^{(\rm eh)}_i} + \sum\nolimits_{j = 1}^{{N^{\rm {id}}}} {{\bm{\mathrm{w}}_j}s^{(\rm id)}_j}.$
${\bm{\mathrm{u}}_i} \in {\mathbb{C}^{{N_t} \times 1}}$ is the energy-bearing signal for the $i$-th EH device with Gaussian distribution, i.e., $\bm{\mathrm{{u}}}_i \sim \mathcal{CN} (\bm{\mathrm{0}},\bm{\mathrm{U}}_i \succeq \bm{\mathrm{0}})$. ${\bm{\mathrm{w}}_j} \in {\mathbb{C}^{{N_t} \times 1}}$ is the information beamforming vector associated with the $j$-th ID device. $s^{(\rm eh)}_i$ with \begin{small}${\mathbb{E}}\left\{ {\left| s^{(\rm eh)}_i \right|^2}\right\}  = 1$\end{small} and $s^{(\rm id)}_j$ with \begin{small}${\mathbb{E}}\left\{ {\left| s^{(\rm id)}_j \right|^2}\right\}  = 1$\end{small} represent the energy and the information signals for the $i$-th EH device and the $j$-th ID device, respectively.
For the $i$-th EH device, the harvested energy can be given by
\begin{equation}
E_i^{\rm {(eh)}} = \zeta_i \mathrm{Tr}\left((\bm{\mathrm{h}}^{\rm{(eh)}}_{0,i})^H\left({\sum\nolimits_{j = 1}^{{N^{\rm {id}}}}}{\bm{\mathrm{w}}_j}{\bm{\mathrm{w}}^H_j} + \bm{\mathrm{\Lambda}}\right)\bm{\mathrm{h}}^{\rm{(eh)}}_{0,i}\right){T},
\label{EH}
\end{equation}
where $\zeta_i \in (0,1]$ is the energy conversion efficiency and $\bm{\mathrm{\Lambda}} \buildrel \Delta \over = \sum\nolimits_{i = 1}^{{N^{\rm {eh}}}} {\bm{\mathrm{U}}_i}$, denoting the energy transmit covariance matrix.
Let $\text{R}_j$ be the desired information rate of the $j$-th ID device. The achievable information rate must exceed the desired information rate, which is given by
 \begin{equation}
 \begin{small}
{\mathcal{C}}\left({\frac{{{{\left| {(\bm{\mathrm{h}}^{\rm(id)}_{0,j})^H{\bm{\mathrm{w}}_j}} \right|}^2}}}{{\sum\nolimits_{k \ne j}^{{N^{\rm {id}}}} {{{\left| {(\bm{\mathrm{h}}^{\rm(id)}_{0,j})^H{\bm{\mathrm{w}}_k}} \right|}^2}} + I_j^{\rm (eh)} + {B}{{\delta ^2}}}}}\right) \ge {\text{R}_j}, \forall j \in \mathsf{N}^{\rm {id}},
\end{small}
\label{qos}
\end{equation}
where $B$ is the system bandwidth, ${\delta ^2}$ is the noise power spectral density, \textcolor[rgb]{0.00,0.00,0.00}{$I_j^{\rm (eh)} = (\bm{\mathrm{h}}^{\rm(id)}_{0,j})^H\bm{\mathrm{\Lambda}}\bm{\mathrm{h}}^{\rm(id)}_{0,j}$, and ${\mathcal{C}}(x) \buildrel \Delta \over = B\log(1+x)$.}

\subsection{UL Model}
In the UL, EH devices offload partial data to the F-HAP with frequency division multiple access (FDMA). Assume each computation task is data partitioned, so it can be divided into two independent parts. Let $D_i$ denote the computation tasks data size of the $i$-th EH device and $O_i$ be the part for fog computing. Thus, the rest data with size of $(D_i - O_i)$ is for local computing.
To complete fog computing, $O_i$ should be offloaded to the F-HAP, so
 \begin{equation}
{\alpha _i}B{t_u}\log \left(1 + \frac{{p_i{{\left\| {{\bm{\mathrm{h}}^{\rm{(eh)}}_{i,0}}} \right\|}^2}}}{{{\alpha _i}B{{\delta ^2}}}}\right) = O_i,
\label{rfog}
\end{equation}
where ${\alpha _i} \in [0,1]$ is the bandwidth allocation factor with $\sum\nolimits_{i = 1}^{N^{\rm {eh}}} {{\alpha _i}}  \le 1$, ${t_u}$ is the offloading time for EH devices, and $p_i$ is the transmit power for offloading. Following (\ref{rfog}), $p_i$ can be expressed by
${p_i} = {{\alpha _i}}\mathscr{F}(\frac{{{O_i}}}{{{\alpha _i}}})/{{{{\left\| {{\bm{\mathrm{h}}^{\rm{(eh)}}_{i,0}}} \right\|}^2}}}$, where ${\mathscr{F}}(x) = {\delta ^2}B({2^{\frac{x}{{B{t_u}}}}} - 1)$.
To complete local computing, the required energy is given by
$$E_{i}^{\rm {(loc)}} = {\kappa _i}(f_i^{\rm {(loc)}})^3{T} = {\kappa _i}\frac{{q_i^3{{\left( {{D_i} - {O_i}} \right)}^3}}}{{{T^2}}},$$
where $\kappa_i$ describes the effective capacitance coefficient that depends on the chip architecture. $f_i^{\rm {(loc)}}$ is the local central processing unit (CPU) frequency represented by ${{q_i{{\left( {{D_i} - {O_i}} \right)}}}}/{{{T}}}$ \textcolor[rgb]{0.00,0.00,0.00}{\cite{F. Wang}} with $q_i$ being the required CPU cycle numbers for computing per bit at the $i$-th device (in cycles/bit).
Since the available energy for computation offloading and local computing is limited by the harvested energy in the DL, it satisfies that
\begin{equation}
E_{i}^{\rm {(loc)}} + {p_i}{t_u} + E_c \le E_i^{\rm {(eh)}}, \forall i \in \mathsf{N}^{\rm {eh}}.
\label{Rewec}
\end{equation}
\textcolor[rgb]{0.00,0.00,0.00}{After offloading, the computation tasks are processed at the F-HAP with delay $(T - {t_u})$, which satisfies that}
 \begin{equation}
T - t_u  \ge \frac{{\sum\nolimits_{i = 1}^{{N^{\rm {eh}}}} {{O_i}{q_i}}}}{{F}},
\label{dataamount}
\end{equation}
where $F$ is computation capacity of F-HAP. For notation simplification, we define $C_{\rm {th}} = F({T- t_u})$ in the sequel.

\subsection{System Design Target}
For such a SWIPT-aware fog computing system, we desire to minimize the required energy at the F-HAP by jointly optimizing energy and information beamforming designs at the F-HAP in the DL, the bandwidth allocation and the computation offloading distribution in the UL. \textcolor[rgb]{0.00,0.00,0.00}{Firstly, the FOT design is studied in order to find some basic insights for the system in Section \ref{FOT}. Then, the OOT design is studied for achieving a better system performance in Section \ref{OOT}.}

\section{The Optimal FOT Design}\label{FOT}
Let $\overline{\bm{\mathrm{w}}} = [\bm{\mathrm{w}}_1^T,..., \bm{\mathrm{w}}_{N^{\rm {id}}}^T]^{T}$ being the total information beamforming vector, $\bm{\mathrm{\alpha}} = [\alpha_1,..., \alpha_{N^{\rm {eh}}}]^{T}$ being bandwidth allocation vector, and $\bm{\mathrm{O}} = [O_1,..., O_{N^{\rm {eh}}}]^{T}$ being computation offloading distribution vector. The energy minimization problem for the optimal FOT design is given by
\begin{subequations} \label{eq0}
\begin{align}
(\mathbb{P}_0) \quad \mathop {\min }\limits_{\overline{\bm{\mathrm{w}}},{\bm{\mathrm{\Lambda}}}, \bm{\mathrm{\alpha}}, \bm{\mathrm{O}}} \quad & \sum\nolimits_{j = 1}^{{N^{\rm {id}}}} {{{\left\| {{\bm{\mathrm{w}}_j}} \right\|}^2}}T + {\mathrm{Tr}({\bm{\mathrm{\Lambda}}})}T + \beta \sum\nolimits_{i = 1}^{{N^{\rm {eh}}}} {{O_i}}\nonumber\\
\text{s.t.} \quad
& (\ref{qos}), (\ref{Rewec}), (\ref{dataamount}), \bm{\mathrm{\Lambda}} \succeq \bm{\mathrm{0}}, \nonumber \\
& 0 \le {\alpha _i} \le 1, \forall i \in \mathsf{N}^{\rm {eh}}, \label{eq0b}\\
& \sum\nolimits_{i = 1}^{N^{\rm {eh}}} {{\alpha _i}}  \le 1,\label{eq0c}\\
& 0 \le {O _i} \le D_i, \forall i \in \mathsf{N}^{\rm {eh}}, \label{eq0d}
\end{align}
\end{subequations}
where $\beta$ describes the energy consumption per bit at the F-HAP (in joule/bit). \textcolor[rgb]{0.00,0.00,0.00}{In the objective function of Problem $(\mathbb{P}_0)$, both transmitting energy and computation energy are considered \textcolor[rgb]{0.00,0.00,0.00}{\cite{F. Wang}}}. Problem $(\mathbb{P}_0)$ is non-convex due to the non-convex constraints (\ref{qos}) and (\ref{Rewec}). To handle the non-convexity, new matrix variables $\bm{\mathrm{W}}_j \buildrel \Delta \over = \bm{\mathrm{w}}_j\bm{\mathrm{w}}_j^H, $ denoting with $\mathrm{rank}(\bm{\mathrm{W}}_{j}) = 1$ are defined.
Then, (\ref{qos}) is rewritten by
\begin{equation}
 \begin{aligned}
\begin{small}\left( {\mathrm{Tr}\big({\bm{\mathrm{G}}_j}\big(\sum\limits_{k \ne j}^{{N^{\rm {id}}}}{\bm{\mathrm{W}}_k}} + {\bm{\mathrm{\Lambda}}}\big)\big) + B{\delta^2}\right)\gamma_j
\le \mathrm{Tr}\left({\bm{\mathrm{G}}_j}{\bm{\mathrm{W}}_j}\right), \end{small} \forall j \in \mathsf{N}^{\rm {id}},
 \end{aligned}
\label{qos1}
\end{equation}
where \begin{small}$\gamma_j = ( {{2^{\tfrac{{{R_j}}}{B}}} - 1} )$\end{small} and ${\bm{\mathrm{G}}_j} = {\bm{\mathrm{h}}^{\rm{(id)}}_{0,j}}{(\bm{\mathrm{h}}^{\rm{(id)}}_{0,j})^H}$.
Moreover, (\ref{Rewec}) is re-expressed as
\begin{equation}
 \begin{aligned}
& {\kappa _i}\frac{{q_i^3{{\left( {{D_i} - {O_i}} \right)}^3}}}{{{T^2}}} + {{{\alpha _i}}}\mathscr{F}(\frac{{{O_i}}}{{{\alpha _i}}})/{{{{\left\| {{\bm{\mathrm{h}}^{\rm{(eh)}}_{i,0}}} \right\|}^2}}}  + E_c \\
& \begin{small} \le \zeta_i \left(\sum\nolimits_{j = 1}^{{N^{\rm {id}}}}\mathrm{Tr}\left({\bm{\mathrm{H}}_{0,i}}{\bm{\mathrm{W}}_j}\right)+ {\mathrm{Tr}\left({\bm{\mathrm{H}}_{0,i}}{\bm{\mathrm{\Lambda}}}\right)}\right){T},\end{small} \forall i \in \mathsf{N}^{\rm {eh}},
 \end{aligned}
\label{ec1}
\end{equation}
where ${\bm{\mathrm{H}}_{0,i}} = {\bm{\mathrm{h}}^{\rm{(eh)}}_{0,i}}{(\bm{\mathrm{h}}^{\rm{(eh)}}_{0,i})^H}$.
Since ${{{\alpha _i}{t_u}}}{\mathscr{F}}(\frac{{{O_i}}}{{{\alpha _i}}})/{{{{\left\| {{\bm{\mathrm{h}}^{\rm{(eh)}}_{i,0}}} \right\|}^2}}} = {\alpha _i}B{\delta^2}({2^{\frac{{{O_i}}}{{{\alpha _i}Bt_u}}}} - 1){t_u}/{{{{\left\| {{\bm{\mathrm{h}}^{\rm{(eh)}}_{i,0}}} \right\|}^2}}}$, which is convex for joint ${\alpha _i}$ and $O_i$, both (\ref{qos1}) and (\ref{ec1}) are convex now.
By introducing $\overline{\bm{\mathrm{W}}} = [\bm{\mathrm{W}}_1,...,\bm{\mathrm{W}}_{N^{\rm {id}}}]$, we adopt SDR to relax Problem $(\mathbb{P}_0)$ to be Problem $(\mathbb{P}_1)$ \cite{XA}, which is given by
\begin{subequations} \label{eq1}
\begin{align}
(\mathbb{P}_1) \quad \mathop {\min }\limits_{\overline{\bm{\mathrm{W}}}, {\bm{\mathrm{\Lambda}}}, \bm{\mathrm{\alpha}}, \bm{\mathrm{O}}}  \quad & f_0 = \sum\nolimits_{j = 1}^{{N^{\rm {id}}}} \mathrm{Tr}\left( \bm{\mathrm{W}}_j\right) T + {\mathrm{Tr}({\bm{\mathrm{\Lambda}}})}T + E_{\rm {cp}}\nonumber\\
\text{s.t.} \quad
& (\ref{dataamount}),(\ref{eq0b}) - (\ref{eq0d}),(\ref{qos1}), (\ref{ec1}),\nonumber \\
& \bm{\mathrm{\Lambda}} \succeq \bm{\mathrm{0}}, \bm{\mathrm{W}}_j  \succeq \bm{\mathrm{0}}, \forall j \in \mathsf{N}^{\rm {id}}, \label{eq1b}
\end{align}
\end{subequations}
where \begin{small}$E_{\rm {cp}} = \beta \sum\nolimits_{i = 1}^{{N^{\rm {eh}}}} {{O_i}}$\end{small}. Problem $(\mathbb{P}_1)$ is convex, which can be solved by known methods, e.g., interior point method.

\newtheorem{pro}{\underline{Proposition}}
\begin{pro}
\label{Prop1}
The optimal solution $\overline{\bm{\mathrm{W}}}^*$ to Problem $(\mathbb{P}_1)$ always satisfies $\mathrm{rank}(\bm{\mathrm{W}}^*_{j}) = 1, \forall j \in \mathsf{N}^{\rm {id}}$.
\end{pro}

\begin{IEEEproof}
The proof is shown in Appendix A.
\end{IEEEproof}
Proposition \ref{Prop1} indicates that the optimal $\bm{\mathrm{w}}^*_{j}$, to Problem $(\mathbb{P}_0)$ can be obtained by eigenvalue decomposition of $\bm{\mathrm{W}}^*_{j}$.

In order to provide more insights of computation offloading, we present the semi-closed forms as shown in Proposition \ref{Prop2} by using the dual decomposition method.

\begin{pro}
\label{Prop2}
For any given dual variables $\{{\bm{\mathrm{\nu}}},\bm{\mathrm{\mu}}\}$, the optimal solution $\bm{\mathrm{\alpha}}^{\diamond}$ and $\bm{\mathrm{O}}^{\diamond}$ to problem $(\mathbb{P}_1)$ satisfies that\\
1) When $\bm{\mathrm{\mu }} = \bm{\mathrm{0}}$, none of EH devices offloads the tasks to the F-HAP, i.e., $\bm{\mathrm{\alpha}}^{\diamond} = \bm{\mathrm{O}}^{\diamond} = \bm{\mathrm{0}}$;\\
2) When $\bm{\mathrm{\mu }} \succeq \bm{\mathrm{0}}$, for the $i$-th EH device with $\mu_i > 0$, it has
$$\begin{small} O_i^{\diamond} = {\left[ {{D_i} - \sqrt {(\beta  + \nu _1{q_i} + {2^{\frac{{\varphi_i^{\diamond}}}{{Bt_u}}}}{{\ln 2{{\delta ^2}}\mu _i}}/{{{{\left\| {{\bm{\mathrm{h}}^{\rm{(eh)}}_{i,0}}} \right\|}^2}}})\frac{{{T^2}}}{{3\mu _i{\kappa _i}q_i^3}}} } \right]^ + } \end{small}$$ and $\begin{small} {\alpha _i^{\diamond}} = \frac{{{O_i^{\diamond}}}}{{{\varphi_i^{\diamond}}}}\end{small},$
where $\varphi_i^{\diamond}$ is expressed by (\ref{so1}),
and for $\mu_i = 0$, it has ${O}_i^{\diamond} = 0,$ and ${\alpha}_i^{\diamond}= 0$.
\end{pro}

\begin{IEEEproof}
The proof is shown in Appendix B.
\end{IEEEproof}

From Proposition \ref{Prop2}, for any given $\{{\bm{\mathrm{\nu}}},\bm{\mathrm{\mu}}\}$, the optimal $\{\bm{\mathrm{\alpha}}^{\diamond},\bm{\mathrm{O}}^{\diamond}\}$ to problem (\ref{eqDfb}) in Appendix B is obtained. Once the optimal $\{{\bm{\mathrm{\nu}}}^*, \bm{\mathrm{\lambda}}^*, \bm{\mathrm{\mu}}^*\} $ is achieved by using some known methods, e.g., subgradient method, the optimal $\{\bm{\mathrm{\alpha}}^*, \bm{\mathrm{O}}^*\}$ can be derived. Then, by plugging $\bm{\mathrm{\alpha}}^*$ and $\bm{\mathrm{O}}^*$ into Problem $(\mathbb{P}_1)$, the optimal $\overline{\bm{\mathrm{W}}}^*$ and $\bm{\mathrm{\Lambda}}^*$ are obtained.

\section{The Optimal OOT Design}\label{OOT}
Considering that the offloading time $t_u$ also have influence on the system performances, we jointly optimize $t_u$ with the variables of Problem $(\mathbb{P}_1)$ in this section.
When $t_u$ becomes a variable, ${{{\alpha _i}{t_u}}}{\mathscr{F}}(\frac{{{O_i}}}{{{\alpha _i}}})/{{{{\left\| {{\bm{\mathrm{h}}^{\rm{(eh)}}_{i,0}}} \right\|}^2}}}$ becomes non-convex w.r.t. $({t_u}, {O_i}, {\alpha _i})$ in (\ref{ec1}). To deal with the non-convex constraint, firstly, we introduce the auxiliary variable $\tilde{\bm{\mathrm{a}}} = {t_u}\bm{\mathrm{\alpha}}$ with $\tilde{a}_i = {t_u}{\alpha _i}, \forall i \in \mathsf{N}^{\rm {eh}}$. Then, the energy minimization problem of the OOT design can be given by
 \begin{subequations} \label{eqQ1}
\begin{align}
(\mathbb{Q}_1) & \quad  \mathop {\min \quad} \limits_{ \overline{\bm{\mathrm{W}}},  \bm{\mathrm{\Lambda}}, \bm{\mathrm{\alpha}},
 \bm{\mathrm{O}}, {t_u}, \tilde{\bm{\mathrm{a}}}}  \quad f_0 \nonumber\\
\text{s.t.} \quad
& (\ref{dataamount}),(\ref{eq0b}) - (\ref{eq0d}),(\ref{qos1}),(\ref{eq1b}),\nonumber \\
& {\kappa _i}\frac{{q_i^3{{\left( {{D_i} - {O_i}} \right)}^3}}}{{{T^2}}} + ({2^{\frac{{{O_i}}}{{{\tilde{a}_i}B}}}} - 1){{\tilde{a}_i}B{{\delta ^2}}}/{{{{\left\| {{\bm{\mathrm{h}}^{\rm{(eh)}}_{i,0}}} \right\|}^2}}} + E_c\label{eqQ1b}\\
& \begin{small} \le \zeta_i \left(\sum\nolimits_{j = 1}^{{N^{\rm {id}}}}\mathrm{Tr}\left({\bm{\mathrm{H}}_{0,i}}{\bm{\mathrm{W}}_j}\right)+ {\mathrm{Tr}\left({\bm{\mathrm{H}}_{0,i}}{\bm{\mathrm{\Lambda}}}\right)}\right){T},\end{small} \forall i \in \mathsf{N}^{\rm {eh}},  \nonumber \\
& 0 \le t_u \le T,  \label{eqQ1c}\\
& {t_u}{\alpha _i} - \tilde{a}_i = 0, \forall i \in \mathsf{N}^{\rm {eh}}.  \label{eqQ1d}
\end{align}
\end{subequations}
\textcolor[rgb]{0.00,0.00,0.00}{With the non-convex equality constraints in Problem $(\mathbb{Q}_1)$, the solution method proposed in Section \ref{FOT} cannot be applied anymore, so a PDD-based method is designed to find the suboptimal solution  to Problem $(\mathbb{Q}_1)$ \textcolor[rgb]{0.00,0.00,0.00}{\cite{R. Guo}}, which consists of two layers, where the inner layer solves the augmented Lagrangian (AL) problem and the outer layer updates the penalty parameter or the dual variables. }


To apply the PDD-based method to solve Problem $(\mathbb{Q}_1)$ with the coupling equalities in (\ref{eqQ1d}), Problem $(\mathbb{Q}_1)$ is firstly transformed into the AL problem form, i.e.,
\begin{equation*}
\begin{aligned}
(\mathbb{Q}_2) & \quad  \mathop {\min }\limits_{\scriptstyle \overline{\bm{\mathrm{W}}}, \bm{\mathrm{\Lambda}}, \bm{\mathrm{\alpha}}, \hfill\atop
\scriptstyle \bm{\mathrm{O}}, {t_u}, \tilde{\bm{\mathrm{a}}}\hfill}  \quad q = f_0 + \sum\nolimits_{i = 1}^{{N^{\rm {eh}}}}\frac{{1}}{{2c}}{{( {\alpha _i}{t_u} - \tilde{a}_i  + c\tilde{\lambda}_i)^2}} \nonumber\\
\text{s.t.} \quad
& (\ref{dataamount}),(\ref{eq0b})-(\ref{eq0d}),(\ref{qos1}),(\ref{eq1b}),(\ref{eqQ1b}),(\ref{eqQ1c}),\nonumber \\
\end{aligned}
\end{equation*}
where $c$ is the penalty parameter and $\tilde{\lambda}_i$ is the dual variable associated with constraint (\ref{eqQ1d}). $\tilde{\bm{\mathrm{\lambda}}}$ is defined as a vector to collect all $\tilde{\lambda}_i$, i.e., $\tilde{\bm{\mathrm{\lambda}}} \buildrel \Delta \over = [\tilde{\lambda}_1,..., \tilde{\lambda}_{N^{\rm {eh}}}]^{T}$. Note that when $c \to 0$, Problem $(\mathbb{Q}_2)$ is equivalent with Problem $(\mathbb{Q}_1)$ \textcolor[rgb]{0.00,0.00,0.00}{\cite{R. Guo}}. Then, we find the minimum of Problem ($\mathbb{Q}_2$) by an iteration process. \textcolor[rgb]{0.00,0.00,0.00}{One can observe that Problem ($\mathbb{Q}_2$) is with a non-convex objective function and a group of convex constraints. The convex constraints can be divided into two independent sets, i.e, set $\mathcal{C}_1$ and set $\mathcal{C}_2$, with separated variables. That is $\mathcal{C}_1$ including (\ref{eq0b})-(\ref{eq0c}) is only associated with $\{\bm{\mathrm{\alpha}}\}$ and $\mathcal{C}_2$ including (\ref{dataamount}), (\ref{eq0d}), (\ref{qos1}), (\ref{eq1b}), (\ref{eqQ1b})-(\ref{eqQ1c}) is associated with $\bm{\mathrm{V}} \buildrel \Delta \over = \{\overline{\bm{\mathrm{W}}}, \bm{\mathrm{\Lambda}}, \bm{\mathrm{O}}, {t_u}, \tilde{\bm{\mathrm{a}}} \}$. Since $\{\bm{\mathrm{\alpha}}\}$ and $\bm{\mathrm{V}}$ are independent, Problem ($\mathbb{Q}_2$) becomes convex when $\{\bm{\mathrm{\alpha}}\}$ is fixed; on the other hand, when $\bm{\mathrm{V}}$ is fixed, Problem ($\mathbb{Q}_2$) becomes convex.} Thus, with the dual variables and the penalty parameter at the $k$-th iteration denoted as $\{\tilde{\bm{\mathrm{\lambda}}}^{(k)}, c^{(k)}\}$, the block coordinate descent (BCD) method \textcolor[rgb]{0.00,0.00,0.00}{\cite{M. Hong}} with two independent blocks, i.e., $\{\bm{\mathrm{\alpha}}\}$ and $\bm{\mathrm{V}}$, is used to solve the primal variables in Problem ($\mathbb{Q}_2$). Then, the $(k+1)$-th dual variables and the penalty parameter can be updated by the $k$-th ones, which are given  by
\begin{equation} \label{eqDV}
\begin{aligned}
\tilde{\lambda}_i^{(k+1)} = \tilde{\lambda}_i^{(k)} +  \frac{{1}}{{c^{(k)}}}{({t_u}{\alpha _i} - \tilde{a}_i)}, \forall i \in \mathsf{N}^{\rm {eh}},
\end{aligned}
\end{equation}
and
\begin{equation} \label{eqP}
\begin{aligned}
c^{(k+1)} = \theta{c^{(k)}},
\end{aligned}
\end{equation}
where $\theta$ is the iteration step size.
By defining
\begin{equation} \label{eqSC}
\begin{aligned}
\tilde{\epsilon} = \mathop {\max }\nolimits_{\forall i \in \mathsf{N}^{\rm {eh}}} \quad \{\left| {\alpha _i}{t_u} - \tilde{a}_i \right|\}
\end{aligned}
\end{equation}
as the stopping criterion, the proposed PDD algorithm is summarized as Algorithm \ref{AL2}.
Following \textcolor[rgb]{0.00,0.00,0.00}{\cite{R. Guo}}, the proposed PDD-based algorithm converges to a Karush-Kuhn-Tucker (KKT) solution to Problem ($\mathbb{Q}_1$).

\begin{small}
\begin{algorithm}[h]
\caption{The Proposed PDD-based Algorithm}
Initialize $\bm{\mathrm{\alpha}}^{(0,0)} = \bm{\mathrm{I}}^{N^{\rm{eh}}\times 1}$, $q^{(0,0)} = 0$, $\tau^{(0)} = \Delta^{(0)} = \tilde{\lambda}_i^{(0)} = 1$, $c^{(0)} = 0.1$.
Set $n = 0$, $k = 0$, $\varepsilon_1 = 10^{-4}$, and $\varepsilon_2 = 10^{-6}$.\\
 {\bf Repeat}\\
\hspace*{0.05in}  {\bf Repeat}
\begin{algorithmic}[1]
      \State Update $\bm{\mathrm{V}}^{(k,n)}$ when fixing $\bm{\mathrm{\alpha}}^{(k,n)}$.
       \State Update $\bm{\mathrm{\alpha}}^{(k,n+1)}$ and $q^{(k,n+1)}$ with $\bm{\mathrm{V}}^{(k,n)}$.
       \State $n \gets n+1$
\end{algorithmic}
\hspace*{0.05in} {\bf Until} $\left| {q^{(k,n)}- q^{(k,n-1)}} \right| \le \varepsilon_1$\\
\hspace*{0.05in}  Calculate $\tilde{\epsilon}^{(k)}$ by (\ref{eqSC}).\\
\hspace*{0.05in}   {\bf If} $\tilde{\epsilon}^{(k)} \le \Delta^{(k)}$ \\
\hspace*{0.15in} Update $\tilde{\lambda}_i^{(k+1)}$ by (\ref{eqDV}).\\
\hspace*{0.05in}   {\bf else} \\
\hspace*{0.15in}  Update $c^{(k+1)}$ by (\ref{eqP}).\\
\hspace*{0.05in}   {\bf end} \\
\hspace*{0.05in}  Set  $\tau^{(k+1)} = 0.6\tau^{(k)}$, $\Delta^{(k+1)} = (\tau^{(k+1)})^{1/6}$ and $k \gets k+1$.\\
{\bf Until}  $\tilde{\epsilon}^{(k)} \le \varepsilon_2$\\
\label{AL2}
\vspace{-0.15in}
\end{algorithm}
\end{small}

\section{Computational Complexity Analysis}
%
According to \textcolor[rgb]{0.00,0.00,0.00}{\cite{K.-Y. Wang}}, the computational complexity can be analyzed by discussing the number of constraints and the scale of variables. \textcolor[rgb]{0.00,0.00,0.00}{Follow it, we analyze the computational complexity of FOT design and OOT design in this section.}

Firstly, for the FOT design, Problem $(\mathbb{P}_1)$ is with the scale of variables $\widetilde{n_1}$, which is on the order of $\left(({N^{\rm id}}+1)N_t^2 + 2{N^{\rm eh}}\right)$. Since there are $X_1 = 5{N^{\rm eh}} + {N^{\rm id}}+ 2$ linear matrix inequality (LMI) constraints with the size of $1$ and $({N^{\rm id}}+1)$ LMI constraints with the size of $N_t$, \textcolor[rgb]{0.00,0.00,0.00}{to solve Problem $(\mathbb{P}_1)$, the computational complexity is} $O_1 = \mathcal{O}(\sqrt {X_1 + ({N^{\rm id}} + 1){N_t}}(\widetilde{n_1}(X_1 + ({N^{\rm id}} + 1){N_t}^3) + \widetilde{n_1}^2(X_1 + ({N^{\rm id}} + 1){N_t}^2) + \widetilde{n_1}^3) )$.

 \textcolor[rgb]{0.00,0.00,0.00}{For the OOT design, to solve Problem $(\mathbb{Q}_2)$}, there are two layers, where the inner layer is with two blocks. So, by denoting $C_a$ and $C_b$ as the computational complexity of the two blocks, respectively, \textcolor[rgb]{0.00,0.00,0.00}{the computational complexity to solve Problem $(\mathbb{Q}_2)$ is} $O_2 = \mathcal{O}(I_{\rm in}I_{\rm out}(C_a + C_b))$, where $I_{\rm in}$ and $I_{\rm out}$ are the number of iterations for the inner layer and the outer layer, respectively. For block one with $\{\bm{\mathrm{\alpha}}\}$, the scale of variables $\widetilde{n_{a}}$ is on the order of ${N^{\rm eh}}$, and then $C_a = \mathcal{O}(\sqrt {2{N^{\rm eh}} + 1}((2{N^{\rm eh}} + 1)(\widetilde{n_{a}} + \widetilde{n_{a}}^2) + \widetilde{n_{a}}^3))$. For block two with $\bm{\mathrm{V}}$, the scale of variables $\widetilde{n_{b}}$ is on the order of $\left(({N^{\rm id}}+1)N_t^2 + 2{N^{\rm eh}}\right)$, and then $C_b = \mathcal{O}(\sqrt {X_2 + ({N^{\rm id}} + 1){N_t}}(\widetilde{n_{b}}(X_2 + ({N^{\rm id}} + 1){N_t^3}) + \widetilde{n_{b}}^2 (X_2 + ({N^{\rm id}} + 1){N_t^2})+ \widetilde{n_{b}}^3))$, where $X_2 = {3 + 3{N^{\rm eh}} + {N^{\rm id}} }$.

\textcolor[rgb]{0.00,0.00,0.00}{Without loss of generality, we suppose ${N^{\rm id}} = b_1{N_t}$ and ${N^{\rm eh}} = b_2{N_t}$, where $b_1$ and $b_2$ are constants. Then, $O_1$ can be approximated to $O_1 \sim \mathcal{O}\left(\sum\nolimits_{x = 2.5}^{11.5} {(N_t)^x}\right) = \mathcal{O}\left(\left({{N_t}^{12.5}-{N_t}^{2.5}}\right)/\left({N_t-1}\right)\right)$ and $O_2$ can be approximated to $O_2 \sim \mathcal{O}\left(I_{\rm in}I_{\rm out}\left(\sum\nolimits_{x = 1.5}^{11.5} {(N_t)^x}\right)\right) = \mathcal{O}\left(I_{\rm in}I_{\rm out}\left({{N_t}^{12.5}-{N_t}^{1.5}}\right)/\left({N_t-1}\right)\right)$. As a result, $O_1$ and $O_2$ are further approximated to $\mathcal{O}({N_t}^{11.5})$ and $\mathcal{O}(I_{\rm in}I_{\rm out}{N_t}^{11.5})$, respectively.}

\section{Numerical Results}
In the simulations, the following system parameters are set according to \textcolor[rgb]{0.00,0.00,0.00}{\cite{F. Wang, O. Munoz, J. Du}}, where $N_t = 6$, ${N}^ {\rm {id}} = {N}^ {\rm {eh}} = 2$, $\beta = 10^{-4}$ J/bit, $B = 2$ MHz, ${\delta ^2} = -80$ dBm, $F = 4$ GHz, $\zeta = 0.8$, $T = 2$ s, $E_c = 10^{-4}T$ J, and $\theta = 0.1$. For EH devices, $D_i = 10$ Kbits, $\kappa_i = 10^{-24}$, and $q_i = 10^3$ cycles/bit $\forall i \in \mathsf{N}^{\rm {eh}}$. For ID devices, $\gamma_j = 0$ dB, $\forall j \in \mathsf{N}^{\rm {id}}$. The distance between the F-HAP and the $i$-th EH device is randomly selected with $d_i \in [5,10]$ m and the Rician channel model is adopted with Rician factor being $3$. The distance between the F-HAP and the $j$-th ID device is randomly selected with $d_j \in [15, 20]$ m and Rayleigh channel model is considered. Moreover, the path-loss exponent is assumed to be $2$.

\begin{figure}[!t]
\centering
\includegraphics[width=0.41\textwidth]{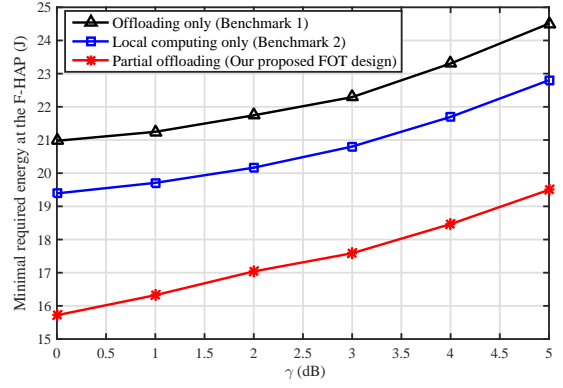}
\caption{The minimal required energy at the F-HAP versus $\gamma$.}
\label{figS1}
\vspace{-3mm}
\end{figure}

\begin{figure}[!t]
\centering
\includegraphics[width=0.41\textwidth]{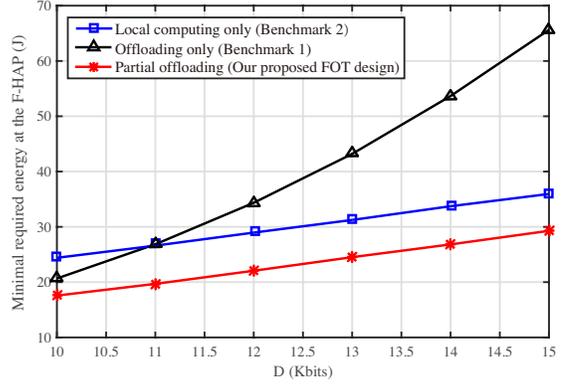}
\caption{The minimal required energy at the F-HAP versus $D$.}
\label{figS2}
\vspace{-3mm}
\end{figure}

For the FOT design, we set $t_u = 0.8T$. Fig. \ref{figS1} and Fig. \ref{figS2} compare the partial offloading mode and two benchmark modes, i.e., ``local computing only'' mode and ``offloading only'' mode, versus $\gamma$ and $D$, respectively. Fig. \ref{figS1} shows the minimal required energy at the F-HAP versus $\gamma$. It is observed that the partial offloading mode is superior to the two benchmark modes. Moreover, the ``offloading only'' mode is worse than the ``local computing only'' mode with the increment of $\gamma$. It means that more sufficient energy supply for EH devices motivates local computing rather than offloading, because local computing requires less energy than computation offloading. Fig. \ref{figS2} depicts the minimal required energy at the F-HAP versus $D$, where the partial offloading mode shows the best performance. One can also see that the ``offloading only'' mode is better than the ``local computing only'' mode with a relatively small $D$ while the ``offloading only'' mode is worse than the ``local computing only'' mode with a relatively large $D$.

\begin{figure}[!t]
\centering
\includegraphics[width=0.41\textwidth]{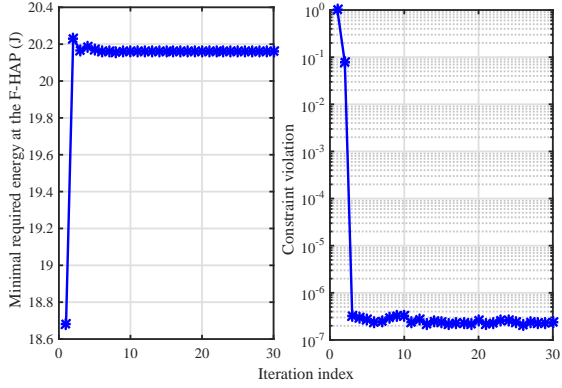}
\caption{The convergency of the PDD-based method with $\gamma = 5$ dB.}
\label{figS3}
\vspace{-3mm}
\end{figure}

For the OOT design, Fig. \ref{figS3} plots the convergency of the PDD-based method with $\gamma = 5$ dB. It is shown that the minimal required energy at the F-HAP converge within 6 iterations for outer iterations of Algorithm \ref{AL2}, and the constraint violation $\tilde{\epsilon}$ reduces to the threshold $10^{-6}$ also in a few iterations for outer iterations of Algorithm \ref{AL2}.

\begin{figure}[!t]
\centering
\includegraphics[width=0.41\textwidth]{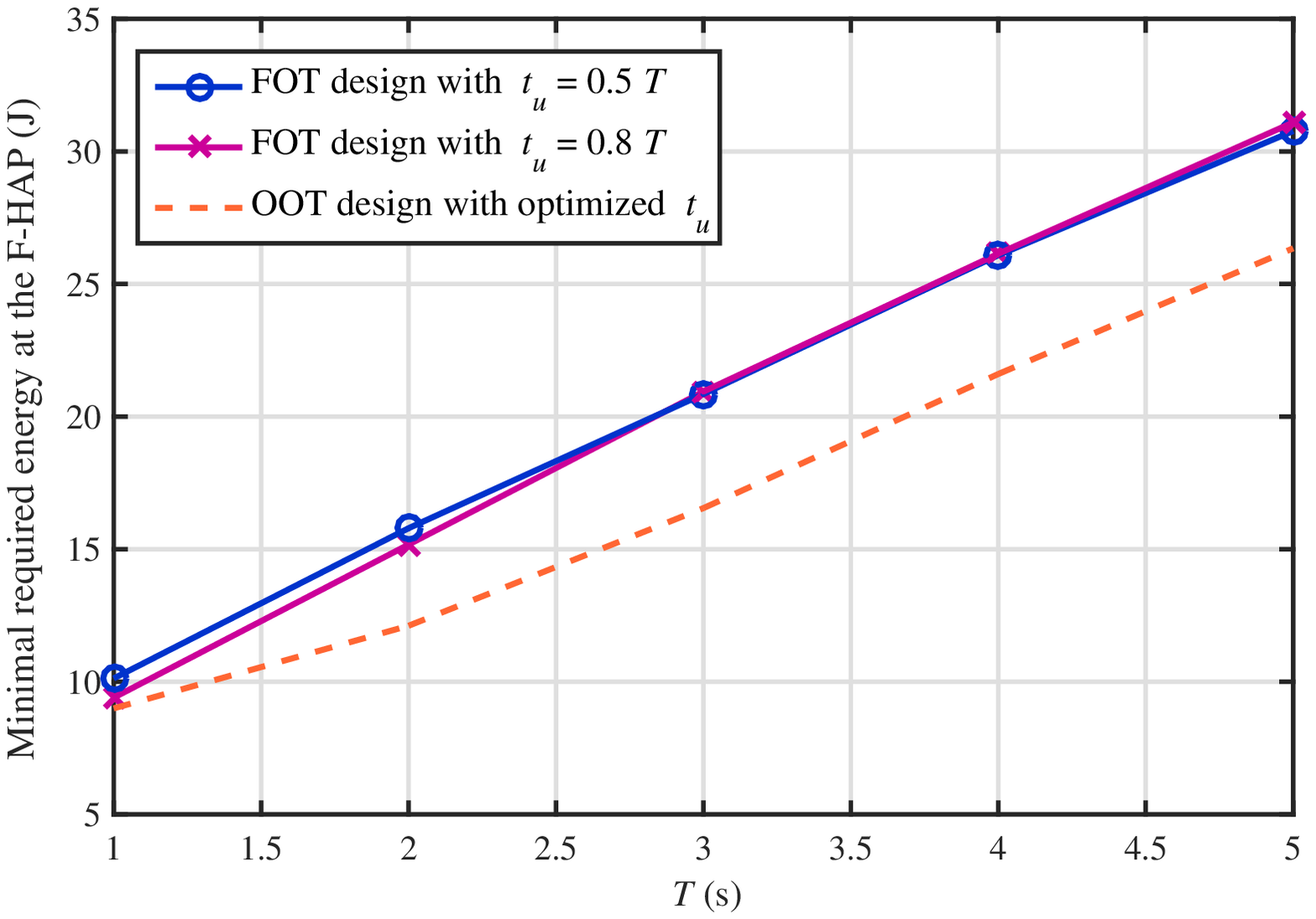}
\caption{The minimal required energy at the F-HAP versus $T$.}
\label{figS4}
\vspace{-2mm}
\end{figure}

\begin{figure}[!t]
\centering
\includegraphics[width=0.41\textwidth]{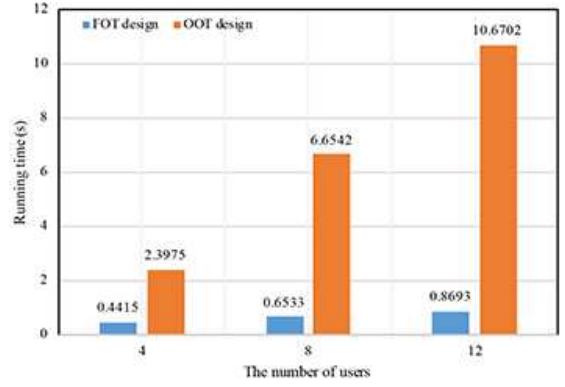}
\caption{The running time versus the number of users.}
\label{figS5}
\vspace{-3mm}
\end{figure}

Fig. \ref{figS4} compares the minimal required energy at the F-HAP versus $T$ with two fixed $t_u$ and the optimized $t_u$ obtained by the PDD-based method. It is seen that by optimizing $t_u$, the required energy can be greatly reduced.
That is the OOT design achieves much better performance than the FOT design. However, as shown in Fig. \ref{figS5}, to achieve such a performance gain, the running time associated with two designs are different. The running time of the OOT design is much higher than that of the FOT design, and the former one increases faster than the latter one with the increment of user numbers.

\section{Conclusions}
This paper studied a SWIPT-aware fog computing network consisting of a F-HAP and multiple heterogeneous IoT devices. An energy consumption minimization problem was formulated by jointly optimizing energy and information beamforming designs at the F-HAP, bandwidth allocation and computation offloading distribution with two designs. For the FOT design, the SDR was adopted and it was proved that the rank-one constraints were always satisfied, the global optimal solution was guaranteed. For the OOT design, since the non-convexity, a PDD-based algorithm was proposed to achieve a suboptimal solution. \textcolor[rgb]{0.00,0.00,0.00}{Simulation results suggest that if the system is with strong enough computing capability, the OOT design is suggested to achieve lower required energy; Otherwise, the FOT design is preferred to achieve a relatively low computation complexity.}

\section*{Acknowledgements}
This work is supported in part by National Key R\&D Program of China (no. 2016YFE0200900),  in part by the General Program of the National Natural Science Foundation of China (NSFC)(no. 61671051), in part by the Fundamental Research Funds for the Central Universities (no. 2017YJS046).

\begin{figure*}
\begin{footnotesize}
\begin{subequations} \label{eqL}
\begin{small}
\begin{align}
& \mathcal{L} =
 {\nu _1}\left(\sum\limits_{i = 1}^{{N^{\rm {eh}}}} {{O_i}{q_i}}  - {C_{\rm {th}}}\right)
 + {\nu _2}\left(\sum\limits_{i = 1}^{N^{\rm {eh}}} {{\alpha _i}}  - 1\right) + \sum\limits_{j = 1}^{{N^{\rm {id}}}} {{\lambda _j}\left(\left({\mathrm{Tr}({\bm{\mathrm{G}}_j}(\sum\limits_{k \ne j}^{{N^{\rm {id}}}}{\bm{\mathrm{W}}_k}} + {\bm{\mathrm{\Lambda}}})) + B{\delta_2}\right)\gamma_j - \mathrm{Tr}({\bm{\mathrm{G}}_j}{\bm{\mathrm{W}}_j})\right) }\nonumber\\
 & + \sum\limits_{i = 1}^{{N^{\rm {eh}}}} {{\mu _i}\left({\kappa _i}\frac{{q_i^3{{\left( {{D_i} - {O_i}} \right)}^3}}}{{{T^2}}} + {\mathscr{F}}(\frac{{{O_i}}}{{{\alpha _i}}})\frac{{{\alpha _i}{t_u}}}{{{{\left\| {{\bm{\mathrm{h}}^{\rm{(eh)}}_{i,0}}} \right\|}^2}}} + E_c  - \zeta_i\mathrm{Tr}\left({\bm{\mathrm{H}}_{0,i}}\left(\sum\limits_{j = 1}^{{N^{\rm {id}}}}{\bm{\mathrm{W}}_j} + \bm{\mathrm{\Lambda}}\right)\right) T\right)}  + \mathrm{Tr}\left( \sum\limits_{j = 1}^{{N^{\rm {id}}}} \bm{\mathrm{W}}_j + \bm{\mathrm{\Lambda}}\right)T + \beta \sum\limits_{i = 1}^{{N^{\rm {eh}}}} {{O_i}},\label{eqL1}\\
 & =
  \underbrace{\sum\nolimits_{j = 1}^{N^{\rm {id}}} \mathrm{Tr}\left( \left(T\bm{\mathrm{I}} - {\lambda_j}{\bm{\mathrm{G}}_j} - \sum\nolimits_{i = 1}^{{N^{\rm {eh}}}} {\mu_i} \zeta_i T{\bm{\mathrm{H}}_{0,i}}+ \sum\nolimits_{k \ne j}^{{N^{\rm {id}}}} {{\lambda _k}{\gamma _k}{\bm{\mathrm{G}}_k}}\right){\bm{\mathrm{W}}_j}\right)}_{(a)} + \underbrace{{\rm{Tr}}\left(\left(\sum\nolimits_{j = 1}^{{N^{\rm {id}}}}{\lambda_j}{\gamma_j}{\bm{\mathrm{G}}_j} -\sum\nolimits_{i = 1}^{{N^{\rm {eh}}}} {\mu_i} \zeta_i T{\bm{\mathrm{H}}_{0,i}}\right){\bm{\mathrm{\Lambda}}}\right)}_{(b)}\nonumber\\
 & +  \underbrace{\sum\nolimits_{i = 1}^{{N^{\rm {eh}}}}\big( \beta{O_i} + \nu_1 {O_i}{q_i} + \nu_2{\alpha_i} + {\mu_i}{{\kappa _i}\frac{{q_i^3{{\left( {{D_i} - {O_i}} \right)}^3}}}{{{T^2}}} + {\mu_i}{{{\alpha _i}{t_u}}}{\mathscr{F}}(\frac{{{O_i}}}{{{\alpha _i}}})/{{{{\left\| {{\bm{\mathrm{h}}^{\rm{(eh)}}_{i,0}}} \right\|}^2}}}} + {\mu_i}E_c \big)}_{(c)} + \underbrace{\sum\nolimits_{j = 1}^{{N^{\rm {id}}}}{\lambda_j}{\gamma_j}B{\delta _2} - \nu_1{C_{\rm {th}}} - \nu_2}_{(d)}\label{eqL2}
\end{align}
\end{small}
\end{subequations}
\end{footnotesize}
\hrulefill
\end{figure*}

\section*{Appendix A: Proof for Proposition \ref{Prop1}}
In order to analyze the optimal $\overline{\bm{\mathrm{W}}}^*$, the Lagrangian function of Problem $(\mathbb{P}_1)$ is rewritten as (\ref{eqL}), where we only consider the the part about $\overline{\bm{\mathrm{W}}}^*$. Hence, according to $(a)$ term in (\ref{eqL2}), a new problem is constructed as
\begin{equation} \label{eqLW}
\begin{small}
\begin{aligned}
\mathop {\max }\limits_{\bm{\mathrm{\lambda}},\bm{\mathrm{\mu}} \succeq \bm{\mathrm{0}}}  \quad \mathop {\min }\limits_{\bm{\mathrm{W}}_j \in {\mathbb{H}^{N_{t}}}} \quad \mathcal{L}_{\bm{\mathrm{W}}_j} &= \sum\nolimits_{j = 1}^{N^{\rm {id}}} \mathrm{Tr}\left({\bm{\mathrm{B}}_j} {\bm{\mathrm{W}}_j}\right) \\
& + \sum\nolimits_{j = 1}^{N^{\rm {id}}}\mathrm{Tr}\left( (- \bm{\mathrm{Z}}_j - {\lambda_j}{\bm{\mathrm{G}}_j})\bm{\mathrm{W}}_j \right) \nonumber
\end{aligned}
\end{small}
\end{equation}
where \begin{small}${\bm{\mathrm{B}}_j} = T\bm{\mathrm{I}}  - \sum\nolimits_{i = 1}^{{N^{\rm {eh}}}} {\mu_i} \zeta_i T{\bm{\mathrm{H}}_{0,i}}+ \sum\nolimits_{k \ne j}^{{N^{\rm {id}}}} {{\lambda _k}{\gamma _k}{\bm{\mathrm{G}}_k}}$\end{small} and $\bm{\mathrm{Z}}_j \succeq \bm{\mathrm{0}}$ is the dual variable associated with (\ref{eq1b}). Since the minimal value of $\mathcal{L}_{\bm{\mathrm{W}}_j}$ cannot be unbounded and  $- \bm{\mathrm{Z}}_j - {\lambda_j}{\bm{\mathrm{G}}_j} \prec \bm{\mathrm{0}}$, ${\bm{\mathrm{B}}_j} $ should be a positive definite matrix with probability one, i.e., $\text{rank}({\bm{\mathrm{B}}_j^*} ) = N_t$.
Then, the KKT conditions of (\ref{eqLW}) associated with $\bm{\mathrm{W}}_j$ are given by
\begin{subequations} \label{eqLWKKT}
\begin{align}
& \bm{\mathrm{Z}}_j^* \succeq \bm{\mathrm{0}}, \lambda_j^* \ge 0,\mu_j^* \ge 0,\label{eqLWKKTa}\\
&  {\bm{\mathrm{Z}}_j^*}{\bm{\mathrm{W}}_j^*} = \bm{\mathrm{0}}, \label{eqLWKKTb}\\
& \bm{\mathrm{Z}}_j^* = {\bm{\mathrm{B}}_j^*} - {\lambda_j^*}{\bm{\mathrm{G}}_j}.\label{eqLWKKTc}
\end{align}
\end{subequations}
From (\ref{eqLWKKTc}), we have
\begin{equation} \label{eqLW1}
\mathrm{rank}(\bm{\mathrm{Z}}_j^*) = \mathrm{rank}({\bm{\mathrm{B}}_j^*} - {\lambda_j^*}{\bm{\mathrm{G}}_j}) \ge {N_t} - 1.
\end{equation}
Moreover, from a basic inequality for the rank of matrices and (\ref{eqLWKKTb}), we have that
\begin{equation} \label{eqLW2}
\begin{aligned}
&\mathrm{rank}({\bm{\mathrm{Z}}_j^*}{\bm{\mathrm{W}}_j^*}) \ge \mathrm{rank}({\bm{\mathrm{Z}}_j^*}) + \mathrm{rank}({\bm{\mathrm{W}}_j^*}) - N_t = 0 \\
&\Rightarrow N_t -  \mathrm{rank}({\bm{\mathrm{Z}}_j^*}) \ge \mathrm{rank}({\bm{\mathrm{W}}_j^*}).
\end{aligned}
\end{equation}
Therefore, based on (\ref{eqLW1}) and (\ref{eqLW2}), it implies that $\mathrm{rank}({\bm{\mathrm{W}}_j^*}) \le 1$. According to (\ref{qos}), ${\bm{\mathrm{W}}_j^*} \ne  \bm{\mathrm{0}}$, so $\mathrm{rank}({\bm{\mathrm{W}}_j^*}) \ne 0$, which implies that $\mathrm{rank}({\bm{\mathrm{W}}_j^*}) = 1$.
$\hfill\blacksquare$

\section*{Appendix B: Proof for Proposition \ref{Prop2}}
Following (\ref{eqL2}), for any given $\{{\bm{\mathrm{\nu}}}$, $\bm{\mathrm{\mu}}\}$, we have that
\begin{equation} \label{eqDfb}
\begin{aligned}
\mathop {\min }\limits_{0 \le {\alpha _i} \le 1,0 \le {O_i} \le {D_i},  \forall i \in \mathsf{N}^{\rm {eh}}}  \quad & \mathcal{L}_{f_c}
\end{aligned}
\end{equation}
where \begin{small}$\mathcal{L}_{f_c}$\end{small} is the (c) term in (\ref{eqL2}).
When ${\mu _i} = 0, \forall i$, $\mathcal{L}_{f_b}$ becomes $\sum\nolimits_{i = 1}^{{N^{\rm {eh}}}}\left( \beta{O_i} + \nu_1 {O_i}{q_i} + \nu_2{\alpha_i}\right)$. Hence, the optimal solution to problem (\ref{eqDfb}) satisfies $\bm{\mathrm{\alpha}}^{\diamond} = \bm{\mathrm{O}}^{\diamond} = \bm{\mathrm{0}}$.
However, when ${\mu _i}>0$, the KKT conditions are listed as
\begin{subequations} \label{eqLfbKKT}
\begin{small}
\begin{align}
& \varpi_i^{(1){\diamond}}, \varpi_i^{(2){\diamond}}, \phi_i^{(1){\diamond}}, \phi_i^{(2){\diamond}} \ge 0,\label{eqLfbWKKTa}\\
&  0 \le {\alpha _i^{\diamond}} \le 1,  0 \le {O _i^{\diamond}} \le D_i,\label{eqLfbWKKTb}\\
& {\alpha _i^{\diamond}} \varpi_i^{(1){\diamond}}, ({\alpha _i^{\diamond}} -1)\varpi_i^{(2){\diamond}},  {O _i^{\diamond}} \phi_i^{(1){\diamond}}, ({O _i^*} - D_i)\phi_i^{(2)*} = 0, \label{eqLfbWKKTc}\\
&\begin{small} {{\nu _2} + \left({\mathscr{F}}(\frac{{{O_i^{\diamond}}}}{{{\alpha _i^{\diamond}}}}) - \frac{{O_i^{\diamond}}}{{{\alpha _i^{\diamond}}}}{{\mathscr{F}}^{'}}(\frac{{{O_i^{\diamond}}}}{{{\alpha _i^{\diamond}}}})\right)\frac{{{\mu _i}t_u}}{{{{\left\| {{\bm{\mathrm{h}}^{\rm{(eh)}}_{i,0}}} \right\|}^2}}} - \phi_i^{(1){\diamond}} + \phi_i^{(2){\diamond}}} = 0,\end{small} \label{eqLfbWKKTd}\\
& {\beta  + {\nu _1}{q_i} - \frac{{3{\mu _i}{\kappa _i}q_i^3{{\left( {{D_i} - {O_i^{\diamond}}} \right)}^2}}}{{{T^2}}} + {{{\mu _i}t_u}}{{\mathscr{F}}^{'}}(\frac{{{O_i^{\diamond}}}}{{{\alpha _i^{\diamond}}}})}/{{{{\left\| {{\bm{\mathrm{h}}^{\rm{(eh)}}_{i,0}}} \right\|}^2}}} \nonumber\\
&- \varpi_i^{(1){\diamond}} + \varpi_i^{(2){\diamond}} = 0,\label{eqLfbWKKTe}
\end{align}
\end{small}
\end{subequations}
where $\varpi_i^{(1){\diamond}}, \varpi_i^{(2){\diamond}}, \phi_i^{(1){\diamond}}, \phi_i^{(2){\diamond}}$ represent the optimal dual variables and ${\mathscr{F}}^{'}(x)$ is the first-order derivative of ${\mathscr{F}}(x)$.
According to (\ref{eqLfbWKKTa})-(\ref{eqLfbWKKTc}), it is derived that $\varpi_i^{(1){\diamond}} = \varpi_i^{(2){\diamond}} = \phi_i^{(1){\diamond}} =  \phi_i^{(2){\diamond}} = 0$. From (\ref{eqLfbWKKTd}), we have that
\begin{equation} \label{so1}
\begin{small}
\begin{aligned}
\varphi_i^{\diamond} \buildrel \Delta \over = \frac{{{O_i^{\diamond}}}}{{{\alpha _i^{\diamond}}}} = \frac{{Bt_u}}{{\ln 2}}\left({W_0}\left({{\nu _2{\left\| {{\bm{\mathrm{h}}^{\rm{(eh)}}_{i,0}}} \right\|}^2}}/({{\mu _i{{\delta ^2}}B{t_u}e}}) - \frac{1}{e}\right) + 1\right),
\end{aligned}
\end{small}
\end{equation}
where $W_0{(x)}$ is the Lambert function \textcolor[rgb]{0.00,0.00,0.00}{\cite{R. Corless}}, and from (\ref{eqLfbWKKTe}), we have
\begin{equation} \label{so2}
\begin{small}
\begin{aligned}
O_i^{\diamond} = {\left[ {{D_i} - \sqrt {(\beta  + \nu _1{q_i} + {2^{\frac{{\varphi_i^{\diamond}}}{{Bt_u}}}}{{\ln 2{{\delta ^2}}\mu _i}}/{{{{\left\| {{\bm{\mathrm{h}}^{\rm{(eh)}}_{i,0}}} \right\|}^2}}})\frac{{{T^2}}}{{3\mu _i{\kappa _i}q_i^3}}} } \right]^ + },
\end{aligned}
\end{small}
\end{equation}
where ${\left[ x \right]^ + } = \max \{ x,0\}$.
Hence, based on (\ref{so1}) and (\ref{so2}), it implies that ${\alpha _i^{\diamond}} = \frac{{{O_i^{\diamond}}}}{{{\varphi_i^{\diamond}}}}$.
$\hfill\blacksquare$

\small
\bibliographystyle{IEEEtran}

\begin{thebibliography}{10}


\bibitem{XB} Q. Wang, D. O. Wu, and P. Fan, ``Delay-constrained optimal link scheduling in wireless sensor networks,"  \emph{IEEE Trans. Veh. Technol.}, vol. 59, no. 9, pp. 4564-4577, Nov. 2010.

\bibitem{XD} K. Xiong, C. Chen, G. Qu, P. Fan, and K. B. Letaief, ``Group cooperation with optimal resource allocation in wireless powered communication networks," \emph{IEEE Trans. Wirel. Commun.}, vol. 16, no. 6, pp. 3840-3853, Jun. 2017.

\bibitem{Y. Lu0} Y. Lu, et al.,``Robust transmit beamforming with artificial redundant signals for secure SWIPT system under non-linear EH model," \emph{IEEE Trans. Wirel. Commun.}, vol. 17, no. 4, pp. 2218-2232, Apr. 2018.

\bibitem{XC} C. Zhang, P. Fan, K. Xiong, and P. Fan, ``Optimal power allocation with delay constraint for signal transmission from a moving train to base stations in high-speed railway scenarios," \emph{IEEE Trans. Veh. Technol.}, vol. 64, no. 12, pp. 5775-5788, Dec. 2015.

%

\bibitem{M. Chiang} M. Chiang and T. Zhang, ``Fog and IoT: an overview of research opportunities,'' \emph{IEEE Internet Things J.}, vol. 3, no. 6, pp. 854-864, Jun. 2016.

\bibitem{C. You} C. You, K. Huang, and H. Chae, ``Energy efficient mobile cloud computing powered by wireless energy transfer,'' \emph{IEEE J. Sel. Areas Commun.}, vol. 34, no. 5, pp. 1757-1771, May 2016.

\bibitem{Y. Mao} Y. Mao, J. Zhang, and K. B. Letaief, ``Dynamic computation offloading for mobile-edge computing with energy harvesting devices,'' \emph{IEEE J. Sel. Areas Commun.}, vol. 34, no. 12, pp.3590-3605, Dec. 2016.


\bibitem{F. Wang} F. Wang, et al., ``Joint offloading and computing optimization in wireless powered mobile-edge computing systems,'' \emph{IEEE Trans. Wireless Commun.}, vol. 17, no. 3, pp. 1784-1797, Mar. 2018.

\bibitem{S. Bi} S. Bi and Y. J. Zhang, ``Computation rate maximization for wireless powered mobile-edge computing with binary computation offloading,"  \emph{IEEE Trans. Wireless Commun.}, vol. 17, no. 6, pp. 4177-4190, Jun. 2018.



\bibitem{N. Janatian} N. Janatian, I. Stupia and L. Vandendorpe, ``Optimal resource allocation in ultra-low power fog-computing SWIPT-based networks," in \emph{Proc. IEEE WCNC}, Barcelona, 2018.

\bibitem{H.N. Zheng} H. N. Zheng, et al., ``SWIPT-aware fog information processing: local computing vs. fog offloading,'' \emph{Sensors}, vol. 18, no. 10, pp. 3291-3307, Sept. 2018.

\bibitem{H. Chai} H. Chai, et al., ``Resources allocation in SWIPT aided fog computing networks," in \emph{Proc. IEEE ICAIT}, Chengdu, pp. 239-244, 2017.


%







\bibitem{XA} K. Xiong, P. Fan, Y. Lu, and K. B. Letaief, ``Energy efficiency with proportional rate fairness in multirelay OFDM networks," \emph{IEEE J. Sel. Areas Commun.}, vol. 34, no. 5, pp. 1431-1447, May 2016.

\bibitem{R. Guo} R. Guo, et al., ``Joint design of beam selection and precoding matrices for mmWave MU-MIMO systems relying on lens antenna arrays,''  \emph{IEEE J. Sel. Topics Signal Process.}, vol. 12, no. 2, pp. 313-325, May 2018.

\bibitem{M. Hong} M. Hong, et al., ``A unified algorithmic framework for block-structured optimization involving big data: With applications in machine learning and signal processing,'' \emph{IEEE Signal Process. Mag.}, vol. 33, no. 1, pp. 57-77, Jan. 2016.

\bibitem{K.-Y. Wang} K.-Y. Wang, et al., ``Outage constrained robust transmit optimization for multiuser MISO downlinks: Tractable approximations by conic optimization,'' \emph{IEEE Trans. Signal Process.}, vol. 62, no. 21, pp. 5690-5705, Nov. 2014.

\bibitem{J. Du} J. Du, et al., ``Computation offloading and resource allocation in mixed fog/cloud computing systems with min-max fairness guarantee," \emph{IEEE Trans. Commun.}, vol. 66, no. 4, pp. 1594-1608, Apr. 2018.

\bibitem{O. Munoz} O. Munoz, et al., ``Optimization of radio and computational resources for energy efficiency in latency-constrained service offloading,'' \emph{IEEE Trans. Veh. Technol.}, vol. 64, no. 10, pp. 4738-4755, 2015.


\bibitem{R. Corless} R. Corless, et al., ``On the Lambert W function,'' \emph{Adv. Comput. Math.}, vol. 5, no. 1, pp. 329-359, Dec. 1996.

\bibitem{XE} Jingyu Kang, Pingyi Fan, and Zhigang Cao, ``Flexible construction of irregular partitioned permutation LDPC codes with low, error floors," \emph{ IEEE Commun. Lett.}, vol. 9, no. 6, pp. 534-536, June 2005.

%


\end{thebibliography}


\end{document}